# Dislocation-induced structural and luminescence degradation in InAs quantum dot emitters on silicon

Eamonn T. Hughes[1], Gunnar Kusch[2], Jennifer Selvidge[1], Bastien Bonef[1], Justin Norman[1], Chen Shang[1], John E. Bowers[1], Rachel A. Oliver[2], Kunal Mukherjee[3,a]

[1]*Materials Department, University of California Santa Barbara, Santa Barbara, California 93106, USA*
[2]*Department of Materials Science and Metallurgy, University of Cambridge, 27 Charles Babbage Road, Cambridge CB3 0FS, United Kingdom*
[3]*Department of Materials Science and Engineering, Stanford University, Stanford, California 94305, USA*

**ABSTRACT**

We probe the extent to which dislocations reduce carrier lifetimes and alter luminescence and growth morphology in InAs quantum dots (QD) grown on silicon. These heterostructures are key ingredients to achieving a highly reliable monolithically integrated light source on silicon necessary for photonic integrated circuits. We find up to 20–30% shorter carrier lifetimes at spatially resolved individual dislocations from both the QD ground and excited states at room temperature using time-resolved cathodoluminescence spectroscopy. These lifetimes are consistent with differences in the intensity measured under steady-state excitation suggesting that trap-assisted recombination limits the minority carrier lifetime, even away from dislocations. Our techniques also reveal the dramatic growth of misfit dislocations in these structures under carrier injection fueled by recombination-enhanced dislocation glide and III-V/Si residual strain. Beyond these direct effects of increased nonradiative recombination, we find the long-range strain field of misfit dislocations deeper in the defect filter layers employed during III-V/Si growth alter the QD growth environment and introduce a crosshatch-like variation in the QD emission color and intensity when the filter layer is positioned close to the QD emitter layer. Sessile threading dislocations generate even more egregious hillock defects that also reduce emission intensities by altering layer thicknesses, as measured by transmission electron microscopy and atom probe tomography. Our work presents a more complete picture of the impacts of dislocations relevant for the development of light sources for scalable silicon photonic integrated circuits.

[a] Corresponding author: kunalm@stanford.edu

# I. INTRODUCTION

Understanding how dislocations affect the properties of optoelectronic devices like lasers and photodetectors is central to efforts in direct epitaxial heterogeneous integration of active devices for silicon photonics.[1] In practice, getting lasers to operate efficiently and reliably in the presence of dislocations remains a key challenge in III-V/Si integration to serve the growing communications market at scale.[2] While advances in experimental and computational methods continue to uncover structure-property relationships of dislocation-induced electronic states and nonradiative recombination of charge carriers in semiconductors like Ge,[3] GaAs,[4] and GaN,[5,6] there is a growing realization that the impact of dislocations goes beyond this static and often idealized picture. Dislocations may affect a heterostructure device even before device operation by altering the local composition or growth rates during synthesis, exemplified by prior work on dislocation-induced phase separation in alloys[7,8] and roughening surfaces.[9,10] Dislocations continue to modify device behavior long after fabrication by diffusing or transporting dopants and other impurities during device operation[11,12] or, more dramatically, by damaging devices via recombination-enhanced dislocation motion where dislocations inject point defects and subsequently increase in length over time via dislocation climb.[13–16]

Understanding these broader impacts of dislocations will further the development of self-assembled epitaxial InAs quantum dot (QD) lasers on silicon.[17–20] These are the most dislocation tolerant datacom-band lasers directly grown on silicon, but we need to continue to improve reliability at high current and high temperatures as well as improve manufacturability and uniformity. Both tasks necessitate a detailed representation of dislocation behavior and their local environment. One important consideration is the direct impact of dislocations on QD formation as the epitaxial growth window for QDs is narrow, and hence more sensitive to perturbations than conventional III-V quantum well (QW) heterostructures. Additionally, nonradiative recombination of charge carriers at dislocations in QD systems remains to be fully characterized. Finally, InAs QD devices retain a sizeable thermal strain due to the silicon substrate that continues to drive recombination-enhanced dislocation motion during operation.[21]

In this work, we use a combination of microanalysis techniques on a model shallow (near to the growth surface) layer of InAs QDs on Si to show that dislocations not only reduce excess carrier lifetimes and emission intensities at room temperature, but they also introduce non-trivial crosshatch- and hillock-induced compositional shifts that locally alter the QD energy levels and intensity. Properly accounting for these effects in laser design and growth can yield improved laser performance and reliability. Our work also anticipates the complex effects of dislocations in the next generation of III-V lasers grown directly on silicon beyond datacom wavelengths such as recent works in the visible[22] and mid-infrared.[23]

# II. EXPERIMENTAL METHODS AND SAMPLE DETAILS

The InAs QD model structure investigated here was previously reported in a multi-modal characterization study.[21] Briefly, we use molecular beam epitaxy (MBE) to synthesize the structure depicted in Fig. 1a with an active layer consisting of a single shallow InAs QD layer embedded in a 7 nm $In_{0.15}Ga_{0.85}As$ quantum well and capped by a 100 nm thick GaAs layer. The QD layer is not intentionally doped. The active layer is grown on a GaAs-on-Si template used for an earlier generation of QD lasers. The template consists of two separate defect-filter structures—a 200 nm thick continuous InGaAs layer and a 10-period strained-layer superlattice of 10nm/10nm $In_{0.1}Ga_{0.9}As$/GaAs (see Fig. 5e for a cross-sectional scanning transmission electron microscopy (STEM) image). The threading dislocation density in the sample is $7\times10^6$ cm$^{-2}$, and the InAs QD density is approximately $5\times10^{10}$ cm$^{-2}$. We also label the locations of misfit dislocation networks in Fig. 1a, which will be relevant for later analysis. The growth conditions (temperature, V/III ratio, growth rate) of the various layers have been described previously.[21]

Optical characterization on the nanoscale was performed by cathodoluminescence spectroscopy (CL). The CL measurements were carried out in an Attolight Allalin 4027 Chronos dedicated CL scanning electron microscope (SEM). CL hyperspectral maps were recorded with an Andor Kymera 328i spectrometer with a focal length of 328 mm, a 150-lines-per-mm grating blazed at 1250 nm, and an Andor 512 px InGaAs diode array camera. Time-resolved CL measurements were performed by triggering the electron gun with the third harmonic of a Nd:YAG laser (355 nm) with a frequency of 80.6 MHz and a pulse width of 7 ps. All CL time decay curves were recorded with a time-correlated single photon counting (TCSPC) setup resulting in a time resolution of about 100 ps. All CL measurements were performed at room temperature with an acceleration voltage of 6 kV (interaction region is ~75 nm radius sphere tangential to sample surface) and a beam current of 30 nA for continuous wave measurements and between 15 pA and 90 pA for pulsed measurements.

Atom probe tips were created using an FEI Helios Dualbeam Nanolab 600 focused ion beam (FIB) microscope using standard 30 kV annular milling steps and a 2 kV broad-area polish to form the final tip shape. Tips were evaporated using a Cameca 3000X HR Local Electrode Atom Probe (LEAP) at 40 K with laser pulsing at a 532 nm wavelength, a 200 kHz repetition rate, and a laser pulse energy of 0.20 nJ. TEM foils were prepared using the FEI Helios Dualbeam FIB and imaged using a ThermoFisher Talos in STEM mode using a bright field detector with a collection angle of 17 mrad.

Electron channeling contrast imaging (ECCI) was performed on a ThermoFisher Apreo S SEM using a three-beam g=040 and g=220 channeling condition. Figure 1b shows a plan-view ECCI image of the structure showing numerous long segments of misfit dislocations along with threading dislocations, which together form the subjects of our study. We have previously determined that these misfit dislocations form just below the 7 nm InGaAs QW/GaAs.[21] The origins of these misfit dislocations, which appear in layers grown nominally below the critical thickness for dislocation glide, is also important to contextualize our results. Briefly, these misfit dislocations form not during growth, but after growth as the sample cools due to a combination of: (1) residual tensile strain in the III-V layers due to thermal expansion mismatch with silicon and (2) local pinning of the threading dislocation segment by the InAs QDs.[24] The formation process is illustrated in Fig. 1c where only unpinned threading dislocation segments glide below the QD layer to form misfit dislocations. We have identified these misfit dislocations as being primarily responsible for degradation in early generations of GaAs-based lasers on silicon and, more recently, in InAs quantum dot lasers on silicon where their effects can now be largely mitigated using strained indium-containing trapping layers.[19]

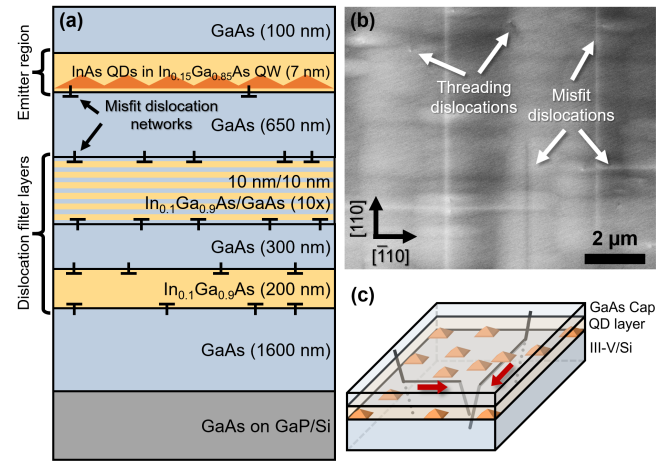

**Fig. 1.** (a) Structure of sample characterized in this study. (b) Electron-channeling contrast imaging (ECCI) from the sample surface showing a moderate density of threading dislocations and misfit dislocations located just below the QD layer. (c) Illustration of the misfit dislocation formation process in which thermal expansion misfit stress generated during cooldown

## III. RESULTS AND DISCUSSION
### A. Recombination dynamics at dislocations

We use time-resolved CL using a pulsed primary electron beam to probe the effect of dislocations on carrier recombination at room temperature. Our results show the misfit dislocations lying close to the InAs QD layers (Fig. 1c) are potent nonradiative recombination sites. Figures 2a-d shows CL intensity decay traces as a function of increasing probe current collected at a dislocation-free region and a region with misfit dislocations. The signal is spectrally filtered to separately track the CL intensity decay of the GS (Fig. 2a-b) and ES (Fig. 2c-d) luminescence at 1250 nm and 1167 nm, respectively, with a 2 nm bandwidth, hence we directly probe only the occupation of dots emitting in these narrow ranges and indirectly probe most remaining dots via their carrier exchange with the wetting layer, due to fast carrier equilibration at room temperature. The insets in these figures show that the recombination lifetime in both regions, obtained by fitting to a single-exponential decay, are in the 0.2–0.3 ns range and do not vary much with probe current. Upon initial inspection, we find the expected outcome that carriers recombine faster near the misfit dislocation, noting a 20% shorter GS recombination lifetime at the lowest probe current. The ES luminescence decays about 30% faster at the misfit dislocation. Figures 2e and 2f show a steady steady-state-excitation CL luminescence map (GS) and a corresponding pulsed-excitation carrier lifetime map obtained from each site. Comparing the two, we see a clear correlation between the CL intensity and luminescence lifetimes, typical of defect-limited recombination.

Figure 2g follows the TRCL decay along a trace that is orthogonal to a misfit dislocation (or group of misfit dislocations) at the center of the distance axis. When carriers are injected directly over the dislocations, nonradiative recombination reduces carrier concentration even at the shortest resolvable time scales (~100 ps, estimated from the signal rise-time), leading to a lowered initial peak intensity at t≈0 s. We may assume that minimal carrier diffusion takes place within this time, so the roughly 1 μm lateral extent of reduced intensity is the convolution of the defect size and the cross section probed by the electron beam. Carriers injected further away from the misfit dislocation should eventually diffuse towards this defect, leading to a widening of the reduced intensity valley with time. Yet, we find that the lateral extent of reduced intensity remains constant even on the longer time scale of 1–2 ns as the luminescence decays, visualized as a trench of apparent constant width in Fig. 2g. Using this information, we obtain an upper bound for the diffusivity of carriers in this system using $L_D = \sqrt{D\tau}$, estimating an ambipolar diffusivity $D$ of less than 40 cm$^2$/s for the measured recombination lifetime $\tau = 0.25$ ns in dislocation-free regions (Fig 2a). This corresponds to a diffusion length, $L_D$, of less than 1 μm, which is shorter than reported values for quantum-well systems in GaAs and reinforces a key mechanism behind the dislocation tolerance of InAs QDs.[25,26] At this time, we are unable to resolve the properties of isolated threading dislocations, but their impact appears minimal compared to misfit dislocations.

We expect these short carrier lifetimes are set by trap-assisted recombination away from dislocations that, naturally, become even shorter at dislocations. Bimberg et al. use PL to measure a spontaneous recombination lifetime, $\tau_r$, in the GS of InAs QDs of 1.8 ns, which is independent of injection over a pulse excitation range of 0.1–100 kW/cm$^2$ at 77 K and only weakly temperature dependent.[27] Fiore et al measure an effective lifetime of 1.8 ns from a single InAs QD layer in an In$_{0.15}$Ga$_{0.85}$As quantum well using PL at room temperature at very low excitation of 9 W/cm$^2$.[28] The much shorter recombination lifetimes measured in our experiments possibly points to elevated point defect concentrations even in regions away from dislocations. Under these constraints, the internal quantum efficiency of spontaneous emission is $\eta = \frac{\tau_{nr}}{\tau_r + \tau_{nr}} \approx \frac{\tau_{nr}}{\tau_r}$, and the recombination lifetime is $\tau = \frac{\tau_{nr}\tau_r}{\tau_{nr}+\tau_r} \approx \tau_{nr}$. Hence, the steady state luminescence of the GS is proportional to the recombination lifetime $\tau$. This is indeed borne out in our experiments where the steady-state GS luminescence peak near misfit dislocations is darker by about 25% (see Section 3.2), comparable to the reduction in lifetime. We see a similar trend for the ES.

In addition to the faster decay at dislocations, there are some features present across the system that are worth noting. Figures 2c-d show a consistently faster decay of the ES intensity compared to the GS both near to and away from dislocations. Dissimilar decay behavior of the ES and GS arise when their occupancy is not in steady state equilibrium with each other and is expected at low temperatures.[29,30] Nevertheless, previous work has shown that the ES and GS start to mirror each other at temperatures above 120 K (for a 60 meV GS-ES energy separation) as the states come into equilibrium with each other.[29] Although our QDs have a slightly larger ES-GS energy separation (about 70 meV), finding dissimilar decay at room temperature is unexpected. Osborne et al. report an anomalous situation in strong electrically pumped InAs dots-in-a-well structure at room temperature where they see the ESs between dots in quasi-equilibrium and the same for the GSs, but unexpectedly, within each dot the ES and GS are not in equilibrium.[31] That is, the ES and GS have different quasi-Fermi energy separations under bias even at room temperature. More work is needed to understand if a similar situation arises in our system that could lead to dissimilar ES and GS decay even at room temperature.

We also note that the GS and ES intensity decay are also slightly non-exponential both near to and away from dislocations as the intensity reduces. Several groups have reported biexponential decay (i.e., a fast and a slow component) of the ES luminescence at cryogenic temperatures.[30,32] In our room-temperature case, it is likely that the origin of non-exponential behavior lies in nonradiative recombination in a disordered system. If dot sizes are inhomogeneous, the dots with deeper confinement lose carriers to traps at a slower rate than shallow dots, once again hinting that global equilibrium is not achieved even at room temperature in these high excitation conditions. We cannot be more definitive about this since our probe directly follows the carrier concentration only over a small range of QD sizes (set by the instrument spectral bandwidth of 2 nm) but still probes other QD sizes indirectly through carrier thermalization and recapture.

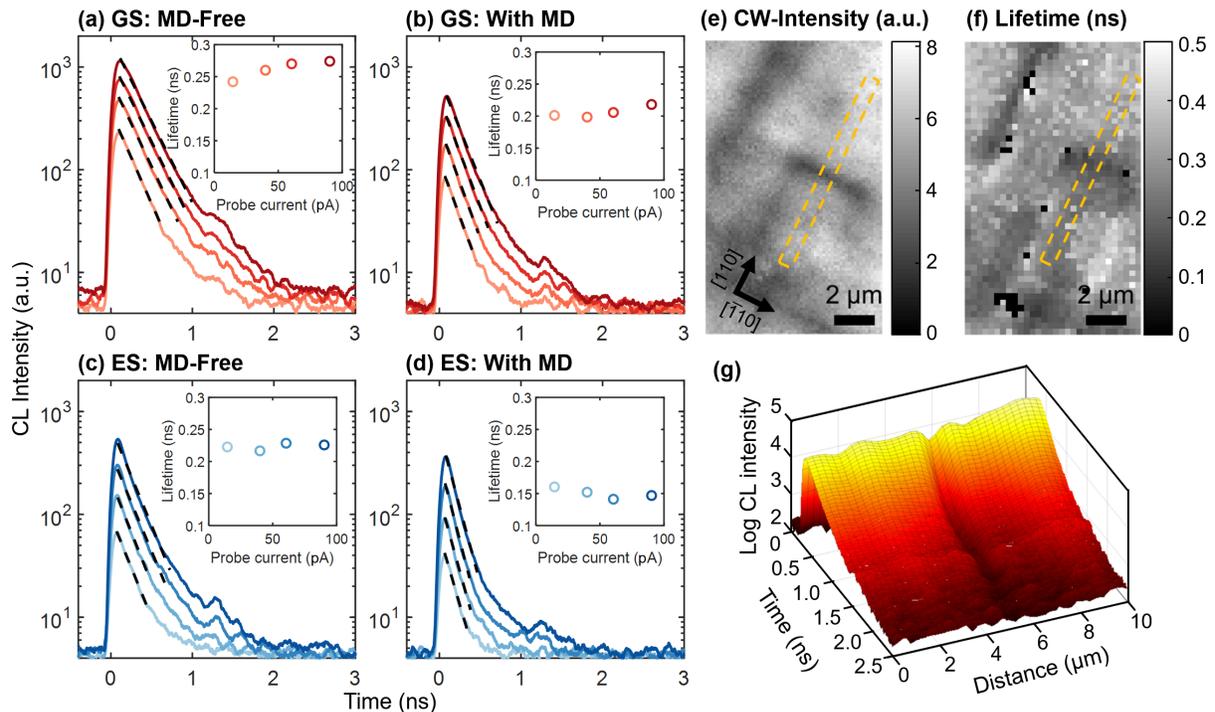

**Fig. 2.** (a-d) Cathodoluminescence intensity decay traces at room temperature as a function of probe current from 15–90 pA for the ground state (a) near to and (b) away from misfit dislocations, and the excited state (c) near to and (d) away from misfit dislocations. The insets show the 1/e lifetimes for each decay trace. (e) Continuous wave cathodoluminescence intensity and (f) 1/e decay lifetime of the same region obtained using a pulsed electron source. The one-to-one correspondence between these two regions demonstrates that nonradiative recombination via dislocation-related traps limits spontaneous emission. (g) Time-position trace of cathodoluminescence intensity across a misfit dislocation (located at 5 μm) taken from the yellow dashed rectangle marked in (e) and (f). A constant width region of reduced intensity corresponding to the misfit dislocation indicates minimal lateral diffusion in the InAs QD system within the experiment window.

## B. In-situ view of recombination-enhanced dislocation glide

The process of nonradiative recombination at dislocations in InAs QDs so far assumes a fixed number of dislocations. However, this is not true in practice. Mismatch in the thermal expansion coefficient of the III-V layers and Si leads to growing tensile strain during cooldown after growth, causing the multi-micron-thick III-V layers to exceed the critical thickness for dislocation glide. While threading dislocations in the epilayers do glide to a certain extent and result in the misfit dislocations characterized earlier, they effectively freeze once temperatures drop below 300 °C, typically leaving a residual strain of about 0.15% at room temperature.

It is now well known that nonradiative carrier recombination at the dislocation core can revive glide even at room temperature via aptly termed recombination-enhanced dislocation glide.[33–35] Figure 3a shows a time-lapse sequence of panchromatic cathodoluminescence (CL) images collected in plan-view, primarily imaging luminescence from the QDs. The sequence of images shows the lengthening of certain misfit dislocation segments along ⟨110⟩ directions after repeated scans. The primary electron beam generates electron-hole pairs that recombine nonradiatively at dislocations and, under the right circumstances, lengthen misfit dislocations by recombination-enhanced dislocation glide. We also see significantly more extension of misfit dislocations along the [$\bar{1}$10] direction over the [110]. In undoped GaAs, α-type dislocation glide is much faster than β- and screw-type dislocations.[36] Thus, we are likely primarily seeing reverse-glide of α-type threading dislocations.[37]

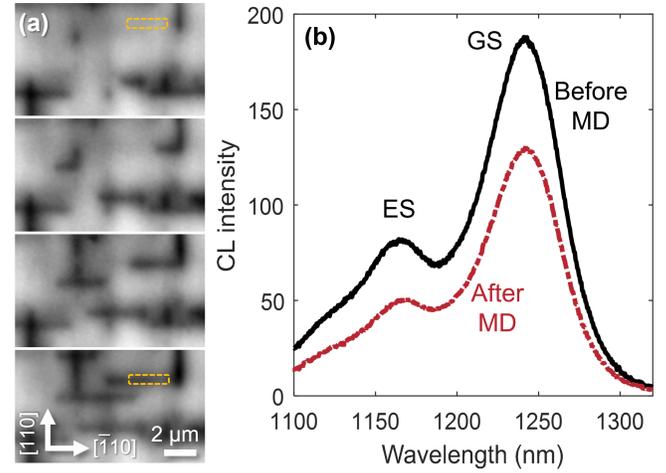

**Fig. 3.** (a) Time-lapse images of recombination-enhanced dislocation glide induced by the scanning electron beam and residual strain in the III-V layer due to thermal expansion mismatch with the silicon substrate. The growing misfit dislocation contrast is captured using panchromatic cathodoluminescence (CL) mapping. The time-lapse was generated from a 6 kV 30 nA scanning electron beam rastered over a 256 um$^2$ area. Each frame in the figure is separated by 30 minutes of scan time. (b) The integrated CL spectra from the yellow dashed rectangle in (a) capture the impact of a misfit dislocation growing.

We probe the impact of the newly grown misfit dislocation in the region marked using the yellow-dotted box (Figure 3a) on QD luminescence in situ. Figure 3b shows luminescence spectra collected over this boxed region before and after the single misfit dislocation grows under it. We measure about a 25% decrease in GS peak luminescence and a 40% decrease in ES luminescence. This difference is reasonable as the lower steady-state carrier concentration near the dislocation implies relatively fewer ES states are filled over GS states. While the newly grown defect reduces the local emission intensity, interestingly, there is no accompanying shift in the luminescence spectrum due to the strong and local strain field of the dislocation. We think this is a consequence of the large interaction volume of the electron beam compared to the extent of the strain field: the dislocation strain field locally affects only a small number of QDs whereas carrier generation, diffusion, and nonradiative recombination affect a much large number of QDs.

## C. Impact of remote misfit dislocations on quantum dot formation

In surveying a wider area of the sample, we find large spatial inhomogeneities in QD emission wavelength and intensity that are distinct from the more local nonradiative effects of dislocations described thus far. Our observation is a potentially important consequence of growth on silicon as the uniformity of emission is key for laser gain and optical isolation. Figure 4a shows a map of the peak GS emission wavelength, respectively, from this sample, exhibiting wide, blue-shifted wavelength bands in a crosshatch-like pattern aligned to the ⟨110⟩ directions. The bands are spaced much wider than the beam interaction cross-section of 100-200 nm diameter convolved with a 1 μm carrier diffusion radius, which points to a long-range effect rather than the typical inhomogeneous broadening from dot-to-dot variation. Each pixel in the map probes luminescence collectively from several hundred QDs (hence already inhomogeneously broadened). A similar sample grown on a GaAs substrate does not exhibit these wide bands of wavelength variation (Fig. 4b), confirming their origin in growth on silicon. Along these blue-shifted bands, the GS emission intensity is also moderately reduced by 10-15% (Fig. 4c). We reiterate that these features are not to be confused with the much more prominent dark regions stemming from the local misfit dislocation network, since, as is clear here and as shown previously in Fig. 3b, these local misfit dislocations are not associated with a wavelength shift. For the sample on GaAs (Fig. 4d), the GS emission intensity is much more uniform, as expected. The corresponding maps for the excited state are shown in Figure S1 and show comparable features to the ground state but with a clearer correlation between blue-shifted bands and reduced emission.

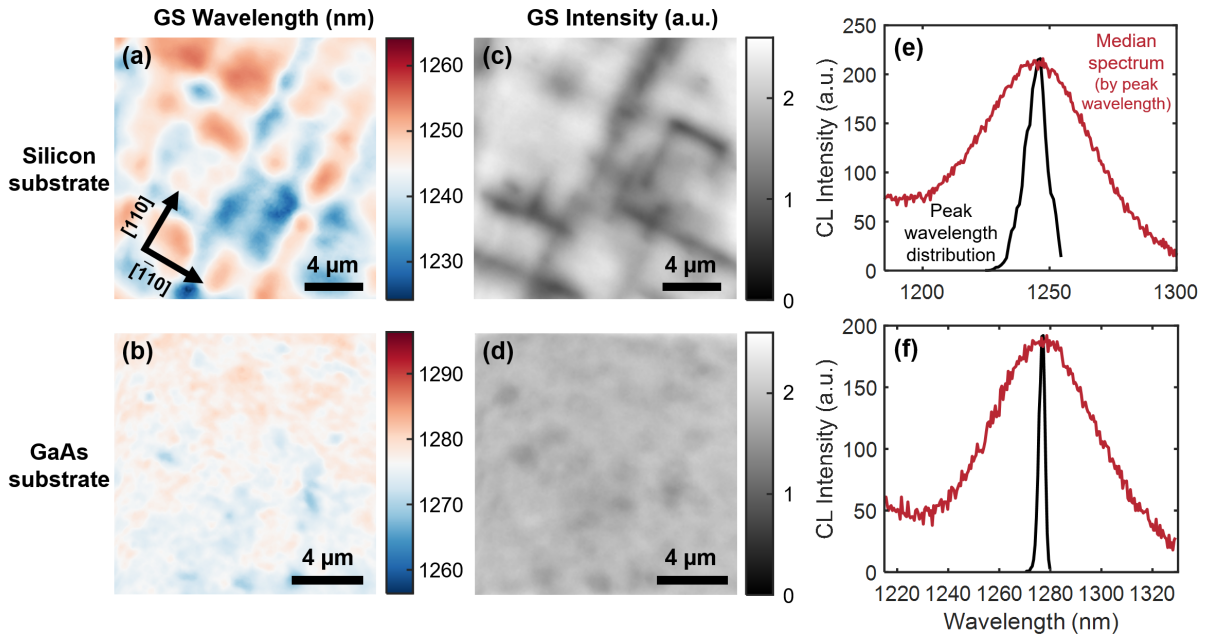

**Fig. 4.** (a-b) Peak emission wavelength of the ground state for InAs QDs grown (a) on silicon and (b) on GaAs collected using steady-state cathodoluminescence hyperspectral imaging. (c-d) Total emission intensity (Gaussian fit) from the ground state (c) on silicon and (d) on GaAs. In addition to sharply reduced intensity at misfit dislocations, a crosshatching in emission intensity and emission wavelength occurs with a reduced intensity in blue-shifted regions. (e) Comparison of a typical pixel spectrum (determined as spectrum with the median GS peak wavelength) (red) to the distribution of peak wavelengths for all spectra in the CL map (black). (f) This same comparison for the sample on GaAs. Comparing (e) and (f), the GaAs sample clearly has a smaller distribution of peak wavelengths; however, both are small compared to the FWHM of the typical spectrum, so overall broadening due to the larger distribution on silicon is muted.

We hypothesize that these darkened, blue-shifted bands arise from the misfit dislocation network lying at the threading dislocation filters layers 650 nm below the QDs, which generates long-range strain fields that alter the growth, and hence emission wavelength and intensity, of the InAs QDs. This points to the important role of dislocation strain fields in influencing the motion of adatoms, particularly indium, during growth and in subtly altering QD formation. The presence of a network of misfit dislocations is known to alter growth rates,[9,38] generate compositional variations in III-V alloy metamorphic layers,[8] and introduce fluctuating surface step densities.[39]

One might expect the significant spatial variation in GS emission seen in CL to be detected by a more routine, spatially unresolved photoluminescence (PL) experiment as a broadened emission peak, but this may often not be the case. We examine the magnitude of this effect in Fig. 4e where we compare the GS peak of a typical pixel to the distribution of all spectra peak wavelengths, weighted by peak intensity. Convolving these two approximately Gaussian distributions gives an approximation of the FWHM when sampling a large area, as is done for typical PL measurements. Despite the significant distribution of peak wavelengths, averaging the spectra over the entire CL map only broadens the FWHM by 1.0 meV or about 2% compared to a typical single pixel FWHM of 44.3 meV. This can be understood by recalling that when convolving two Gaussians, the FWHMs combine as the root of the sum of the squares, so the broadening effect of the relatively tight peak wavelength distribution is greatly suppressed. Comparing to the sample grown on GaAs (Fig. 4f), where these spatial variations are absent, the broadening is negligible with a FWHM of 38.5 meV for both the median pixel and the entire image. While the broadening is certainly larger for the sample on silicon, it is still too small to distinguish from typical sample-to-sample variation. Therefore, spectral measurements made by photoluminescence (PL), a commonly relied upon tool for assessing growth quality, will in many cases be ineffective at detecting this non-uniform crosshatched emission. Further, the associated intensity reduction can also be obscured because PL intensities are generally not comparable between samples and particularly between different substrate types due to differences in reflection at the interface. However, micro-PL mapping with a sufficiently small spot size should be capable of detecting these local wavelength and intensity variations.

Solutions to reduce crosshatch nonuniformity require either reducing adatom diffusivity[8] (by increasing the V/III ratio, for example) or increasing the spacing between the misfit dislocation network and the active layer.[7] During growth of our single-QD-layer sample, the nearest misfit dislocation network lies 650 nm below the QDs at the defect filter layer as shown in Fig. 5e (remember that the other sparse misfit dislocation network adjacent to the QDs only forms later during cooldown). Fortunately, the nearest misfit dislocation network in a typical QD laser is often about twice as distant due to a thick lower AlGaAs cladding. Indeed, we see no crosshatch-like spatial variations (Fig. S2) in a CL map of the active layer from a laser bar, despite a modest density of misfit dislocations formed by post-growth thermal glide. This confirms our hypothesis of long-range strain fields from the buffer as the underlying cause behind crosshatched emission wavelength. Even so, future laser designs, intended to better couple the optical mode from the III-V gain region into silicon and to reduce the likelihood of film cracking call for much thinner buffers and cladding layers.[40–42] If such lasers are directly grown on silicon, the misfit dislocation network may be close enough to the active region to result in undesirable luminescence broadening.

## D. Growth modification near threading dislocations

We have seen that the distant misfit dislocation network influences QD growth itself by altering some combination of the composition, morphology, and thickness of the layers. Yet, the influence of these remote misfit dislocations must be small compared to threading dislocations continuously intersecting the growth surface at a point that may not change much over time. This allows growth impacts to accumulate, in some cases forming growth mounds or hillocks due to locally accelerated growth at spiral step edges. We locate a cluster of threading dislocations shown in Figure 5a using ECCI and place a fiducial marker to co-locate this site in CL and APT. Some threading dislocations appear at the center of hillocks, demonstrating their potential impact on surface morphology. Fig. 5b shows significantly dimmer and blue-shifted emission from the hillock center compared to a region away from the hillock, with no clear GS or ES peaks identified from the former. Fig. 5c shows that the region near the hillock with strongly blue-shifted QD peak emission wavelength overlaps almost exactly with the region of reduced intensity in Fig. 5d. This correspondence likely arises as carriers more easily thermalize out of the GS of the shallower blue-shifted QDs to recombine nonradiatively at the cluster of adjacent threading dislocations. Still, we note that it is primarily the hillock and not the threading dislocations themselves that induce these changes: individual threading dislocations not associated with hillocks elsewhere in the film do not show such blue-shifted emission. Furthermore, it appears that clusters of immobile threading dislocations are required to form hillocks large enough to see these effects, so reducing threading dislocation densities should significantly reduce their incidence.

Hillocks arise due to the spiraling nature of surface steps at threading dislocations that have a screw-component to their Burgers vector. The increased density of step edges surrounding the hillock provides additional nucleation sites for QDs, which may result in a greater number of smaller (in volume), and hence bluer-emitting, QDs. We see tentative evidence for this in cross-sectional STEM of a region containing a threading dislocation with a hillock shown in Fig. 5e. When tilted away from the zone axis, the growth plane containing the QDs at the defect-free region is viewed at an angle in projection (Fig. 5f). When viewing the hillock region in this same tilt condition, the QD growth plane is viewed edge on (Fig. 5g), indicating this growth plane is inclined relative to the zone axis, since this narrow slice of QDs are grown along the side of a hillock. It is also worth considering why hillocks do not feature prominently in conventional III-V lattice-mismatched (metamorphic) growth but do so in our samples. Typically, threading dislocations glide rapidly to relieve strain during growth and tend not to stay in one place long enough to yield a hillock. We speculate that a combination of near on-axis (001) substrate (limiting the density of contending steps) and sessile threading dislocations that arise at the GaAs/Si interface or by dislocation reactions result in hillocks.

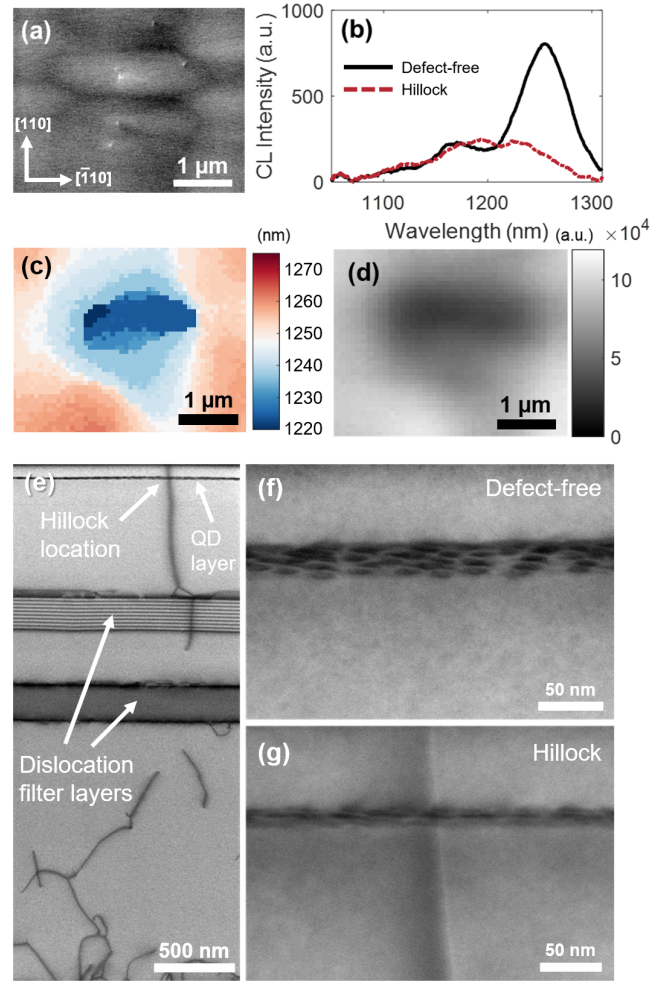

**Fig. 5.** (a) Electron contrast channeling image (ECCI) of a cluster of threading dislocations forming a hillock. (b) Emission spectra from the center and away from the hillock. (c) Peak emission wavelength and (d) peak emission intensity surrounding the hillock region shown in (a). (e) Cross-sectional scanning transmission electron microscopy (STEM) of a region containing a hillock capturing a threading dislocation and a perceived local widening of the active region. High magnification view at (f) a defect-free region showing a low-angle side view of individual InAs QDs due to manual tilting of the foil and (g) the hillock containing a threading dislocation for the same foil tilt, but here, the QDs are viewed edge-on due to compensating tilt of the growth plane surrounding the hillock.

To probe the structural and compositional changes caused by these hillocks in more detail, we extract tips for laser-pulsed APT from the TD-impacted hillock region and from a nominally TD-free region next to it. Note that the shallow 100 nm depth of our QDs that enables CL imaging (and ECCI), dramatically reduces the likelihood of capturing a QD in the APT tip since the conical tip diameter is very small near the top. Indeed, we see from the top-down views in Fig. 6a that neither tip has regions of high indium concentration as would be expected from a QD, indicating that both tips probe only the InGaAs QW that encases the QDs. Nevertheless, the fluctuations are essentially consistent with those of a random alloy of InGaAs, as shown in Fig. S3 and no evidence for phase separation or clustering is seen. On the other hand, the cross-sectional indium profiles of each tip in Fig. 6b reveal that the QW in the defective region is significantly thicker than in the TD-free region. Collapsing these down to one-dimensional vertical profiles of indium composition, averaged laterally over the center of the tip, we see in Fig. 6c that the QW in the hillock is about 8-9 nm thick with a tapering indium profile, contrasted with a 7 nm thick QW with a slightly less tapered profile seen in the tip from the TD-free region. Some of this taper, along with indium concentrations elevated above the expected 15% nominal value, may be explained by the unresolved InAs wetting layer (consistent with other APT[43,44] and STEM[45] studies) that lies 2 nm above the base of the QW. However, the additional thickness of the QW is an effect of the hillock.

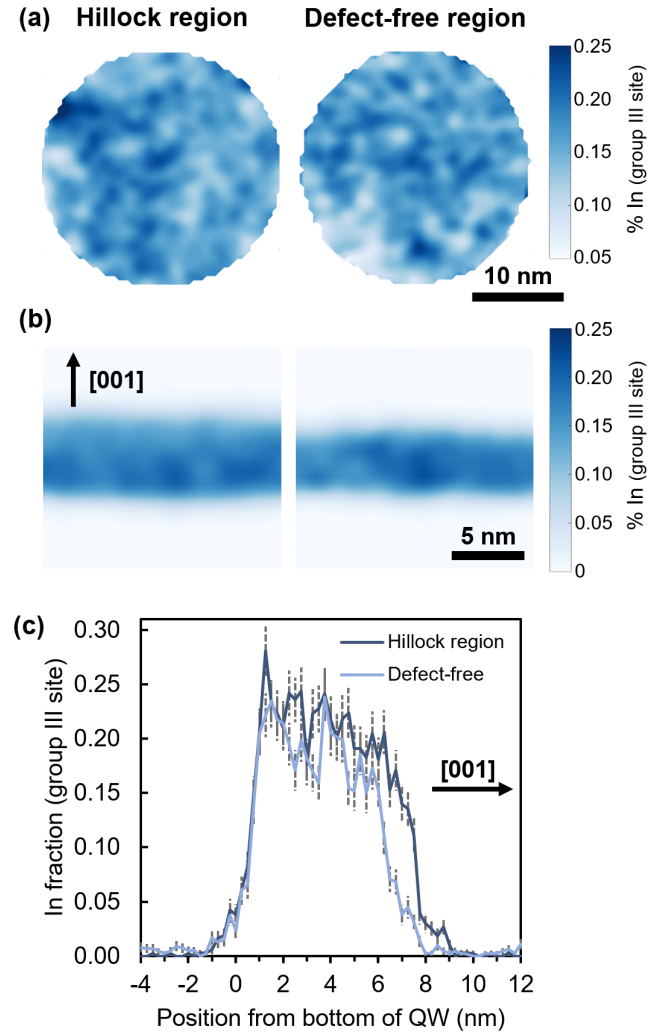

**Fig. 6.** (a) Top view and (b) side view of the lateral indium composition in the nominally $In_{0.15}Ga_{0.85}As$ quantum well that surrounds the InAs quantum dots at a threading dislocation containing hillock, similar to that in Fig. 5 (left) and at a neighboring threading dislocation (TD)-free region (right). Data collected using site-selective laser atom probe tomography informed by cathodoluminescence and electron channeling contrast imaging. No quantum dots were captured in the analysis due to the limited cross-sectional area of the APT tip possible from the 100 nm shallow structure. (c) Vertical composition trace through the quantum well showing a region of tapered but similar composition profiles for the two sites, but increased thickness for the defective region. Error bars representing one standard deviation are indicated by the dotted lines.

Reiterating that the hillock regions contain a higher density of surface steps, if the availability of steps limits the incorporation of adatoms, any asymmetry between the diffusivity of indium and gallium may lead to preferential incorporation of indium in such hillocks. However, the vertical profiles show near identical indium incorporation for the first 7 nm in both sites; the hillock simply extends this tapered profile for an additional 1–2 nm. This suggests that the growth rate increases at the hillock without any alteration in the composition, and both indium and gallium are quite mobile on the growth surface and incorporate at the hillock without preference. Without direct access to the composition or shape of the InAs QDs, we may only infer how the altered QW affects the emission spectra. In addition to easier thermalization from smaller blue-shifted QDs, a locally thicker QW may have a ground state closer in energy to the QDs and enhance carrier thermalization out of the dots. Taken together, these analyses demonstrate the serious impact both distant misfit dislocations and local threading dislocations can have in altering the growth of QDs and their surrounding structures, ultimately broadening their size distribution (and hence their emission spectrum) and further aggravating nonradiative recombination. Therefore, these effects must be closely considered when tuning device design to optimize performance and reliability.

## IV. CONCLUSIONS

With the large untapped potential of heterogeneous integration of dissimilar materials by direct growth, it is important to understand the microscale effect of dislocations on the final devices. We have quantified how dislocations affect spontaneous-emission luminescence in InAs QDs on silicon by facilitating defect-assisted recombination using time-resolved cathodoluminescence spectroscopy on a model InAs QD structure on silicon. We find a significantly reduced recombination lifetime for both the ground and excited states at misfit dislocations but also find recombination to be limited by defects in regions away from dislocations. Yet, the impact of dislocations goes much beyond simple nonradiative recombination. We find, using hyperspectral CL imaging and atom probe tomography, alterations in QD and QW growth that form pockets of blue-shifted emission arising from long range misfit dislocation strain fields and short-range threading dislocation spiral growth. Both yield reduced emission homogeneity that increases susceptibility to carrier losses. Our work shows how new characterization tools may enable a more complete understanding of the impact of dislocations on devices. InAs quantum dots, currently yielding the most reliable devices, are now part of a series of III-V laser devices being synthesized on silicon spanning the visible to the mid-infrared. As the field matures, we expect to see multi-modal microstructural characterization of the kind employed in this work to rise to prominence in those devices as well.

## SUPPLEMENTARY MATERIAL

See supplementary material for (Fig. S1) cathodoluminescence peak excited state wavelength and intensity maps for samples on GaAs and silicon, corresponding to the ground state maps in Fig 4a-d, (Fig. S2) a cathodoluminescence wavelength map of an InAs QD laser active region after milling away the upper cladding, and (Fig. S3) an atom probe compositional frequency distribution comparison of the hillock and defect-free regions.

## ACKNOWLEDGEMENTS

The sample growth was supported by ARPA-E, U.S. Department of Energy, under Award No. DE-AR0001043. This material is based upon work supported by the National Science Foundation (NSF) Graduate Research Fellowship under Grant No. 1650114. APT and TEM studies were performed at the UCSB MRL Shared Experimental Facilities, supported by the MRSEC Program of the NSF under Award No. DMR 1720256; a member of the NSF-funded Materials Research Facilities Network. CL studies were supported by the EPSRC under EP/R025193/1. K.M. acknowledges additional support from NSF CAREER award under grant no. DMR-2036520.

## CONFLICTS OF INTEREST

The authors have no conflicts to disclose.

## DATA AVAILABILITY

The data that support the findings of this study are available from the corresponding author upon reasonable request.

# Supplementary Material

## Dislocation-induced structural and luminescence degradation in InAs quantum dot emitters on silicon

Eamonn T. Hughes[1], Gunnar Kusch[2], Jennifer Selvidge[1], Bastien Bonef[1], Justin Norman[1], Chen Shang[1], John E. Bowers[1], Rachel A. Oliver[2], Kunal Mukherjee[3]

[1]Materials Department, University of California Santa Barbara, Santa Barbara, California 93106, USA
[2]Department of Materials Science and Metallurgy, University of Cambridge, 27 Charles Babbage Road, Cambridge CB3 0FS, United Kingdom
[3]Department of Materials Science and Engineering, Stanford University, Stanford, California 94305, USA

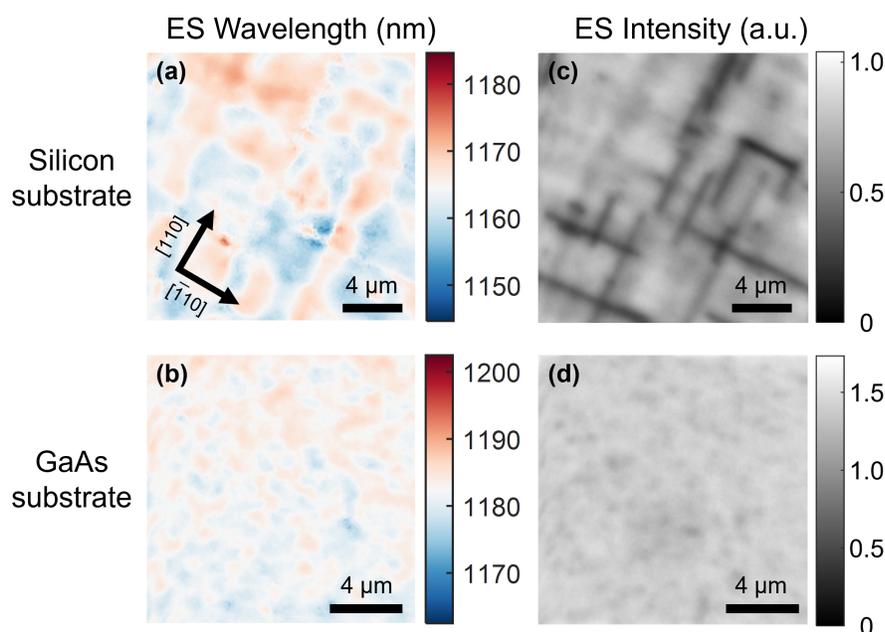

**Figure S1.** (a-b) Excited-state peak-emission wavelength cathodoluminescence map for the sample (a) on silicon and (b) on GaAs. (c-d) Corresponding excited-state cathodoluminescence intensity maps for the sample (a) on silicon and (b) on GaAs.

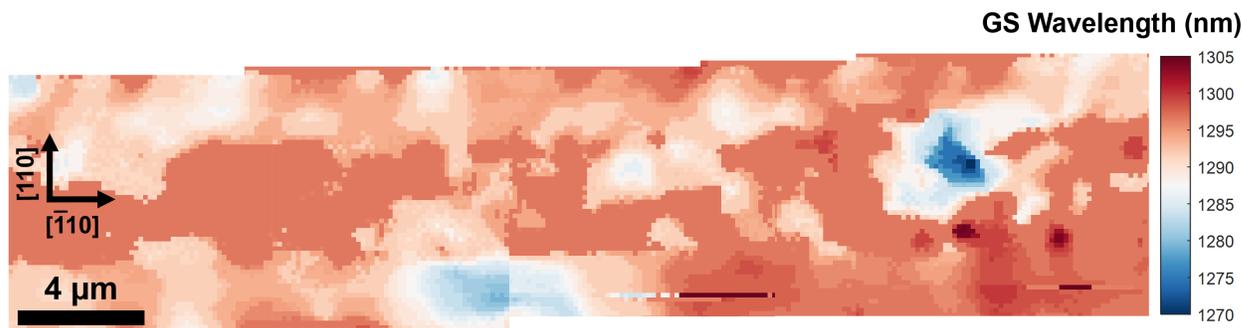

**Figure S2.** Stitched cathodoluminescence map from a five-layer QD laser grown on silicon after milling away upper cladding using a focused ion beam microscope. The spacing between the active region and uppermost defect filter layer (which hosts a misfit dislocation network) is much larger here than in the single QD structure in the main text. Consequently, the effects of extended misfit dislocation strain fields are weaker, so no distinct crosshatch pattern is visible. Even so, there are wide variations in peak emission wavelength and several strongly blue shifted regions, possibly due to hillocks formed by sessile threading dislocation clusters.

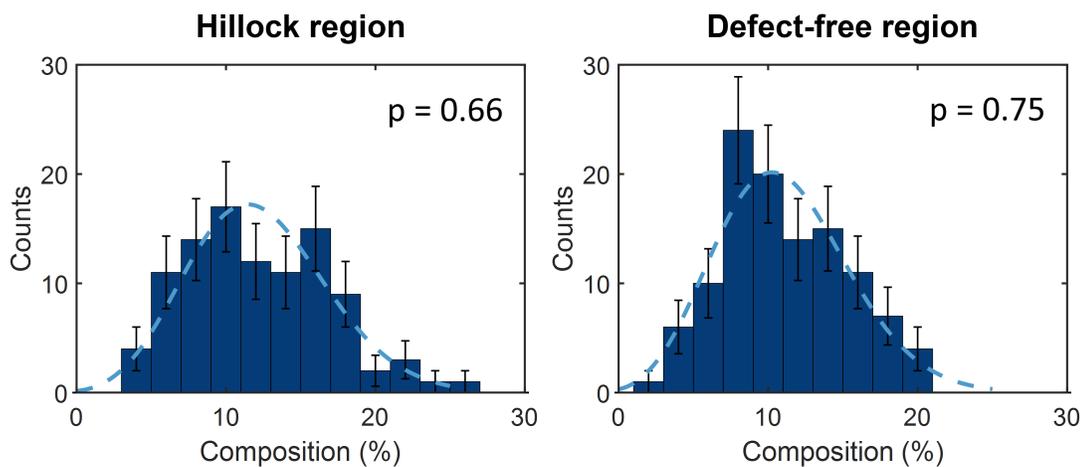

**Figure S3.** Compositional frequency distribution measured from the bottom 2 nm of the QW analyzed in the two atom probe tomography specimens, which roughly aligns with the expected location of any QDs and the wetting layer. The dashed curve is a binomial fit representing the expected compositional distribution for a random alloy. The p-values estimate the probability that the observed distributions represent a random alloy, therefore, both alloys appear to be randomly distributed with no indication of a quantum dot or partial quantum dot present in either. The bin size for composition measurements is 50 atoms.